\documentclass[fleqn,12pt,twoside]{article}
\usepackage{espcrc1}

\usepackage{graphicx}
\usepackage[figuresright]{rotating}

\newcommand{\gtrsim}{\:\lower 0.4ex\hbox{$\stackrel{\scriptstyle >}
{\scriptstyle\sim}$}\:}
\newcommand{\lesssim}{\:\lower 0.4ex\hbox{$\stackrel{\scriptstyle <}
{\scriptstyle\sim}$}\:}

\newcommand{\AmS}{{\protect\the\textfont2
  A\kern-.1667em\lower.5ex\hbox{M}\kern-.125emS}}

\title{Nuclear physics and astrophysics of the $r$-process}

\author{Y.-Z. Qian\address{School of Physics and Astronomy, 
        University of Minnesota, \\ 
        Minneapolis, MN 55455, U.S.A.}%
        \thanks{This work was supported in part by US DOE grants 
	DE-FG02-87ER40328 and DE-FG02-00ER41149.}}
       
\begin{document}

\maketitle

\begin{abstract}
Some nuclear and astrophysical aspects of the $r$-process are
discussed. Particular attention is paid to observations of abundances 
in metal-poor stars and their implications for the astrophysical site 
and yield patterns of the $r$-process. The effects of supernova
neutrinos and related nuclear processes on the yield patterns are
explored. The uncertainties in the theoretical nuclear input for the 
$r$-process are discussed and the need for experimental data is
emphasized.
\end{abstract}

\section{INTRODUCTION}
An overview of the nuclear physics and astrophysics involved in 
$r$-process nucleosynthesis is given here. This is largely a concise 
summary of a recent review on the subject \cite{qian03} by this 
author. An earlier but still very useful review is \cite{cow91}.

When the universe was born 13.5 Gyr ago, the big bang imprinted the
baryonic matter with a primordial composition of 76\% of H and 24\% 
of $^4$He (by mass). By comparison, when the solar system was formed
4.6 Gyr ago, its birth material contained 71\% of H, 27\% of $^4$He,
and 2\% of heavier nuclei. This change in nuclear abundances resulted
from 9 Gyr of nucleosynthesis in stars. Two neutron capture processes
are especially important as they are the dominant mechanisms for
producing the nuclei heavier than the Fe group. These are the slow
($s$) neutron capture process in stars of $\sim 1$--$8\,M_\odot$ and
the rapid ($r$) neutron capture process to be discussed here.
Since the $s$-process operates close to the stable nuclei and its
astrophysical site has been established, its contributions to the
solar system abundances are relatively well understood 
\cite{kap89,arl99}. Consequently, the $r$-process contributions can
be obtained by subtracting the $s$-process contributions from the
total solar abundances. The solar $r$-process (abundance) pattern 
derived this
way is shown in Figure 1. It can be seen that the prominent features 
of this pattern are the peaks at mass numbers $A\sim 130$ and 195.

\begin{figure}[htb]
\begin{minipage}[t]{8cm}
\includegraphics[scale=0.9]{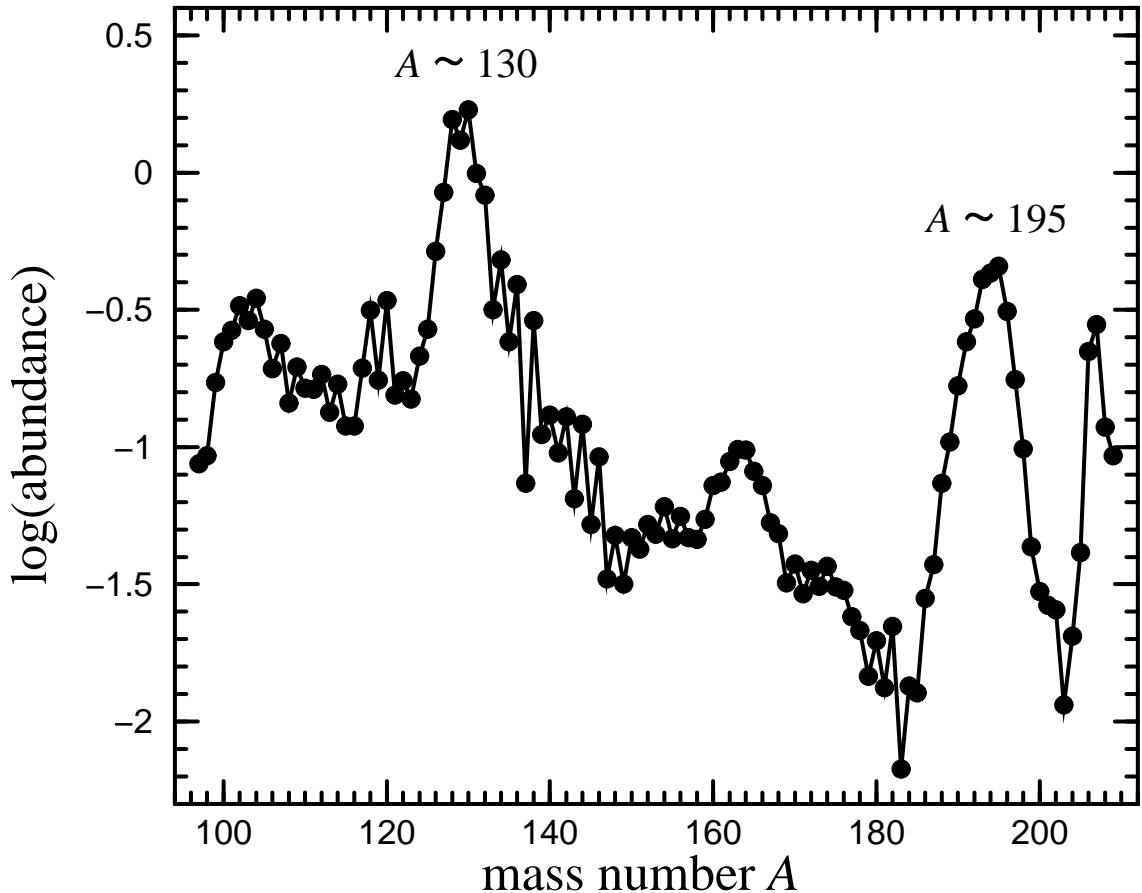}
\end{minipage}
\caption{Solar $r$-process (abundance) pattern as derived in \cite{kap89}.}
\end{figure}

The origin of these peaks can be understood as follows. At the 
beginning of the $r$-process, there are many neutrons and some seed
nuclei to capture them. By definition, neutron capture occurs much
more rapidly than $\beta$ decay during the $r$-process. So the 
$r$-process path moves toward the neutron-drip line as the seed 
nucleus keeps capturing neutrons. However, at some point, the
separation energy of the next neutron to be captured becomes so small
that it will be quickly disintegrated by the photons in the $r$-process
environment. At this so-called waiting point, there is a tug of war 
between neutron capture and photodisintegration. No further net neutron
capture can occur until the waiting-point nucleus $\beta$-decays to a
new species. The tug of war then repeats at the next waiting point.
Through such a series of neutron capture and $\beta$ decay, a population
of very neutron-rich progenitor nuclei far from stability are produced.
Clearly, at a given proton number, the progenitor abundance is piled
at the corresponding waiting-point nucleus. The more slowly this nucleus
$\beta$-decays, the more abundant it will be. The extremely slow 
$\beta$-decay of the progenitor nuclei with magic neutron numbers
$N=82$ and 126 then leads to peaks in the progenitor abundance 
distribution. These become the peaks
at $A\sim 130$ and 195 in the solar $r$-process pattern following
successive $\beta$ decay of the progenitor nuclei upon cessation of
neutron capture.

\section{ASTROPHYSICAL MODELS AND OBSERVATIONS}
As discussed above, given an initial state with many neutrons and
some seed nuclei, simple nuclear systematics will produce the dominant
features of the solar $r$-process pattern. On the other hand, how the
neutrons and seed nuclei are provided must be determined by the 
astrophysical site of the $r$-process. 

One possible site is supernovae 
that occur when a stellar core collapses into a compact neutron star.
The neutron star has an enormous gravitational binding energy of
$\sim 10^{53}$ erg. Due to the high temperature and density encountered 
during the collapse, the most efficient way to release this energy is
to emit $\nu_e$, $\bar\nu_e$, $\nu_\mu$, $\bar\nu_\mu$, $\nu_\tau$, and
$\bar\nu_\tau$. Near the neutron star, the temperature is several MeV
and the material is dissociated into neutrons and protons. As the
neutrinos emitted from the neutron star pass through this material,
some of the $\nu_e$ and $\bar\nu_e$ are captured by the neutrons and
protons, respectively: 
\begin{eqnarray}
\nu_e + n\to p + e^-,\label{nue}\\
\bar\nu_e + p\to n + e^+.\label{nueb}
\end{eqnarray}
The energy deposited by these $\nu_e$ and $\bar\nu_e$ heats the material
and drives a mass outflow usually referred to as the neutrino-driven
wind \cite{dun86}. The reactions in equations (\ref{nue}) and (\ref{nueb})
not only provide the heating to drive the wind but also interconvert
neutrons and protons, thereby determining the neutron-richness of the
wind. Since the $\bar\nu_e$ producing the neutrons have a higher luminosity
and a higher average energy than the $\nu_e$ producing the protons, the
wind is neutron rich (e.g., \cite{qian93,qian96}). As the wind expands
away from the neutron star, its temperature and density decrease and
various nuclear reactions take place to change its composition. Essentially
all the protons are assembled into $\alpha$-particles when the temperature
$T$ drops to $\sim 0.5$ MeV. The material at this $T$ dominantly 
consists of neutrons and $\alpha$-particles. As $T$ drops
further, an $\alpha$-process occurs to burn neutrons and $\alpha$-particles
into heavier nuclei \cite{woo92}. By the time the Coulomb barrier eventually
stops all charged-particle reactions at $T\sim 0.25$ MeV, nuclei with
$A\sim 90$ have been produced. These nuclei then become the seed nuclei to
capture the remaining neutrons during the subsequent $r$-process. This is 
the neutrino-driven wind model of $r$-process nucleosynthesis in supernovae
(e.g., \cite{wob92,mey92,tak94,woo94}).

Another $r$-process model takes advantage of the neutrons inside the
neutron star. When an old neutron star merges with another neutron star 
or a black hole in a binary due to energy loss from gravitational radiation,
some neutron star matter is ejected (e.g., 
\cite{lat74,lat76,ruf97,ros99}). An $r$-process could occur during the
decompression of this extremely neutron-rich ejecta (e.g., 
\cite{mey89,frt99}).

Both the neutrino-driven wind and the neutron star merger models have 
merits. However, due to our insufficient understanding of supernovae and 
neutron star properties, it is difficult to demonstrate ab initio how a 
robust $r$-process occurs in either model. While the conditions required
for an $r$-process have been explored to some detail (e.g., 
\cite{mey97,hof97,fre99}), how these conditions can be realized without
parametrization in the neutrino-driven wind or neutron star mergers
is still an unresolved issue (e.g., \cite{qian96,frt99,tho01}).

\subsection{Supernovae vs. neutron star mergers: observational clues}
Observations of elemental abundances in old Galactic halo stars have
shed important light on the astrophysical site of the $r$-process. 
These stars are very poor in ``metals'' such as Fe since they were formed 
very early in the Galaxy when only a small number of supernovae had 
occurred to provide Fe. With a rate $\sim 10^3$ times lower than
supernovae, neutron star mergers would not have occurred at all for such 
early times. If neutron star mergers were the major source for $r$-process
elements such as Eu, then very metal-poor stars would have no Eu.
This is in strong conflict with the
substantial Eu abundances observed in stars having Fe abundances as low as
$\sim 10^{-3}$ times solar and also with the substantial Ba abundances
(at such low metallicities corresponding to early times,
only the $r$-process associated with fast-evolving massive progenitors
can contribute to the Ba in the interstellar medium) observed in stars 
having Fe abundances as low as $\sim 10^{-4}$ times
solar (e.g., \cite{mcw95,bur00,joh01}).
These observations cannot be accounted for by neutron star mergers being
the major $r$-process site unless the rate of these events is close to that 
of supernovae. Such a high neutron star merger rate is very unlikely and 
it is much more probable that supernovae are responsible for both the
$r$-process and Fe enrichment of metal-poor stars (e.g., 
\cite{qian00,arg04}). In the rest of the discussion, it will be assumed 
that supernovae are the major $r$-process site.

\subsection{Diverse supernova $r$-process sources: more observational clues}
Different conditions are required to produce the peaks at $A\sim 130$ and 
195, respectively, in the solar $r$-process pattern. These peaks may be
produced by the different kinds of supernovae associated with the
neutrino-driven wind model of the $r$-process. The diversity of supernova
sources for the $r$-process was first suggested in \cite{was96} based on
meteoritic data and later received support from observations of the
abundances for a wide range of $r$-process elements in the metal-poor
star CS 22892--052 \cite{sne00}. Further observations of abundances in 
metal-poor stars have also provided important clues in this regard.

\begin{figure}[htb]
\begin{minipage}[t]{8cm}
\includegraphics[scale=0.46]{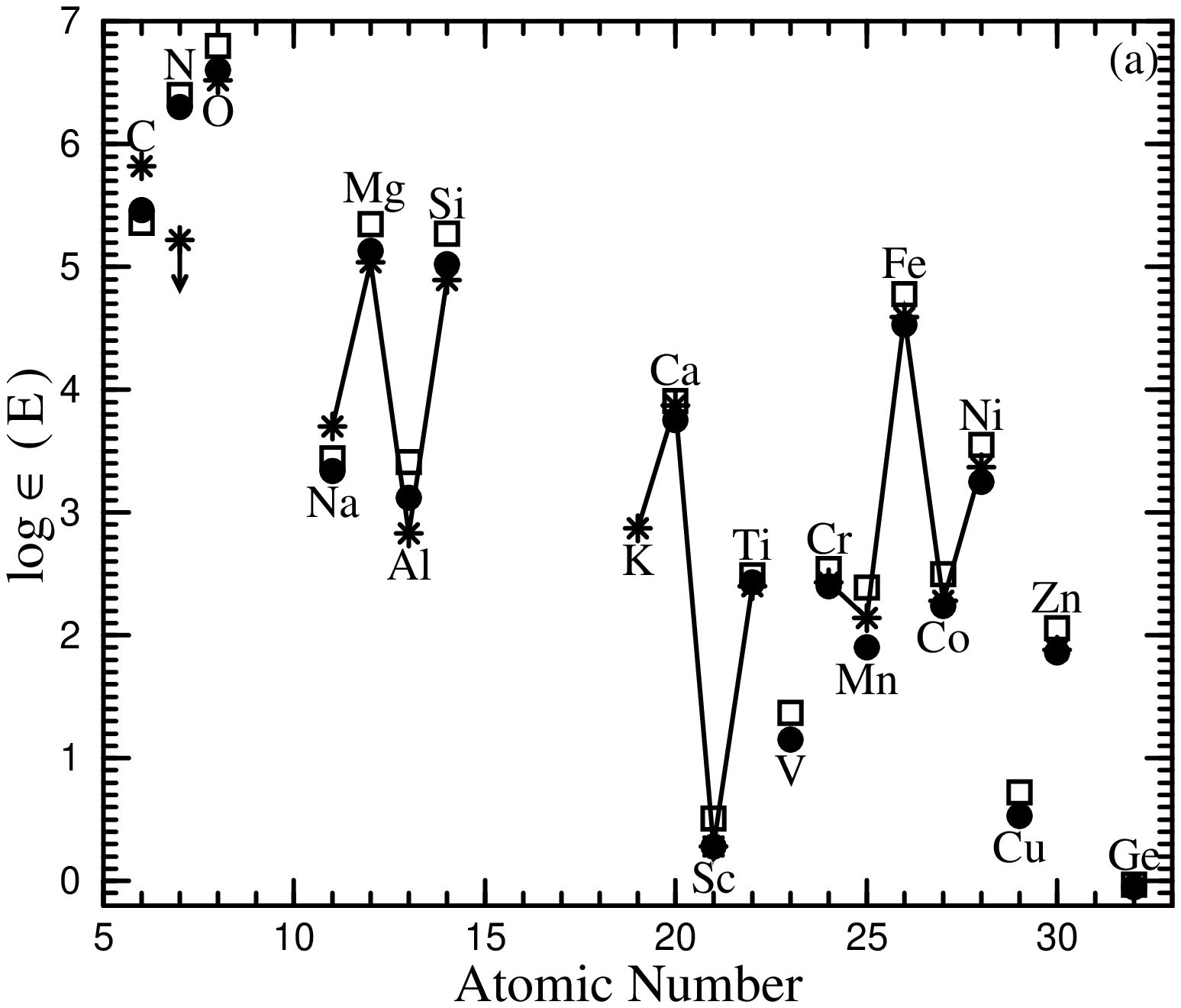}
\end{minipage}
\begin{minipage}[t]{8cm}
\includegraphics[scale=0.46]{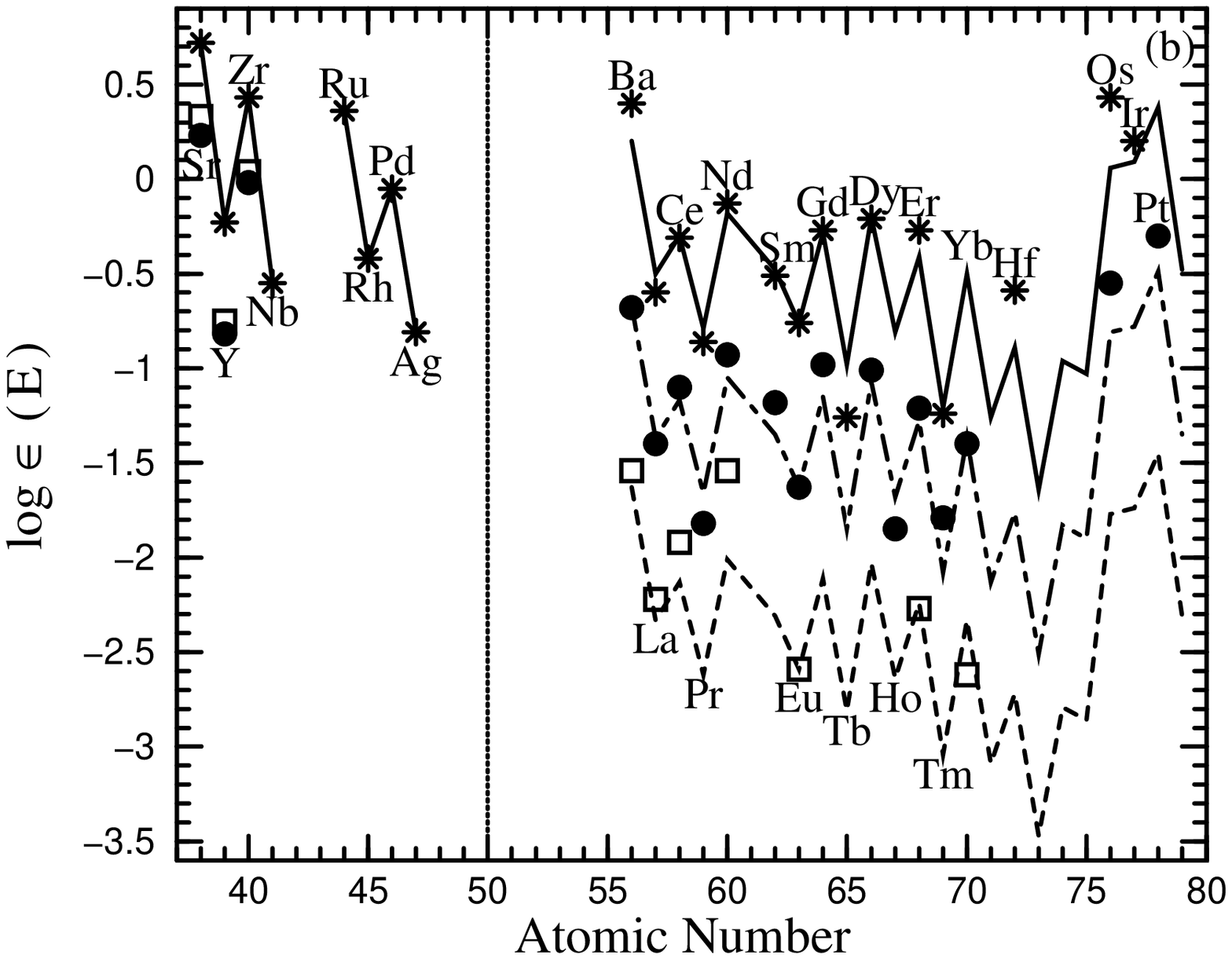}
\end{minipage}
\caption{Comparison of the observed abundances in CS 31082--001 (asterisks:
\cite{hil02}), HD 115444 (filled circles), and HD 122563
(squares: \cite{wes00}). (a) The data on CS 31082--001 are
connected with solid line segments as a guide. Missing segments mean
incomplete data. The downward arrow at the asterisk for N indicates an
upper limit. Note that the abundances of the elements from O to Ge are
almost indistinguishable for the three stars. (b) The data on CS 31082--001
to the left of the vertical line are again connected with solid line
segments as a guide. In the region to the right of the vertical
line, the solid, dot-dashed, and dashed curves are the solar $r$-process
pattern translated to pass through the Eu data for CS 31082--001, HD 115444,
and HD 122563, respectively. Note the close description of the data
by these curves. The shift between the solid and the dashed curves 
corresponds to a change by a factor of $\sim 100$ in the absolute 
abundances.}
\end{figure}

The data on three metal-poor stars CS 31082--001 \cite{hil02}, HD 115444,
and HD 122563 \cite{wes00} are shown in Figure 2. The curves to the right
of the vertical line in Figure 2b represent the solar $r$-process pattern 
translated for comparison with the data on each star. It can be seen that
for all three stars, Ba and heavier elements with $A>130$ (including the
$A\sim 195$ peak at Pt) closely follow
the solar $r$-process pattern. Note that the absolute abundances of these
elements ($\log\epsilon$ is the logarithm of the absolute abundance in 
appropriate units) differ by a factor of $\sim 100$ for CS 31082--001 and
HD 122563. In other words,
if HD 122563 received contributions from one supernovae, then CS 31082--001
received contributions from $\sim 100$ supernovae. However, as shown in
Figure 2a, the abundances of O to Ge including Fe are essentially the same
for these stars. This suggests that supernovae producing the heavy 
$r$-process elements with $A>130$ cannot produce any of the elements from
O to Ge (e.g., \cite{qw02}). Since the latter elements are produced by 
hydrostatic and 
explosive burning in the shells outside the core, to avoid their production
requires the supernovae to have very thin or no shells at all. These
supernovae are associated with progenitors of $\sim 8$--$10\,M_\odot$
\cite{nom84,nom87}
and with accretion-induced collapse of white dwarfs \cite{nom91}. 
On the other hand,
supernovae with progenitors of $>10\,M_\odot$ produce the elements from
O to Ge including Fe (e.g., \cite{woo95}). 
They must then be responsible for the light
$r$-process elements up to the $A\sim 130$ peak. 

\section{NUCLEAR PHYSICS AND YIELD PATTERNS}
As discussed in the introduction, the peaks at $A\sim 130$ and 195
in the solar $r$-process pattern are due to the nuclear systematics 
of the $N=82$ and 126 closed neutron shells, respectively. The 
possible connections between nuclear systematics and features of 
the $r$-process pattern observed in metal-poor stars are discussed below. 
In particular, the effects of reactions induced by supernova neutrinos
are explored considering that the $r$-process occurs in the neutrino-driven 
wind. An example is given to illustrate the uncertainties in the theoretical
nuclear input for the $r$-process and to highlight the need for experimental 
data.

\subsection{Fission and neutrino-induced reactions}
Observations have shown that Ba and heavier elements with $A>130$ in
a number of metal-poor stars closely follow the solar $r$-process pattern
(e.g., Fig. 2b). This highly regular pattern may be produced by some 
special nuclear process. For example, if the $r$-process occurs in an
extremely neutron-rich environment, the heaviest nucleus produced may
fission. The fission fragments then become the new seed nuclei to capture
neutrons. This results in a cyclic flow between the fissioning nucleus and
its fragments. It was shown that for certain conditions, this fission
cycling can produce a solar $r$-process pattern involving only the nuclei 
with $A>130$ (e.g., \cite{frt99}).

\begin{figure}[htb]
\begin{minipage}[t]{8cm}
\includegraphics[scale=0.8,angle=270]{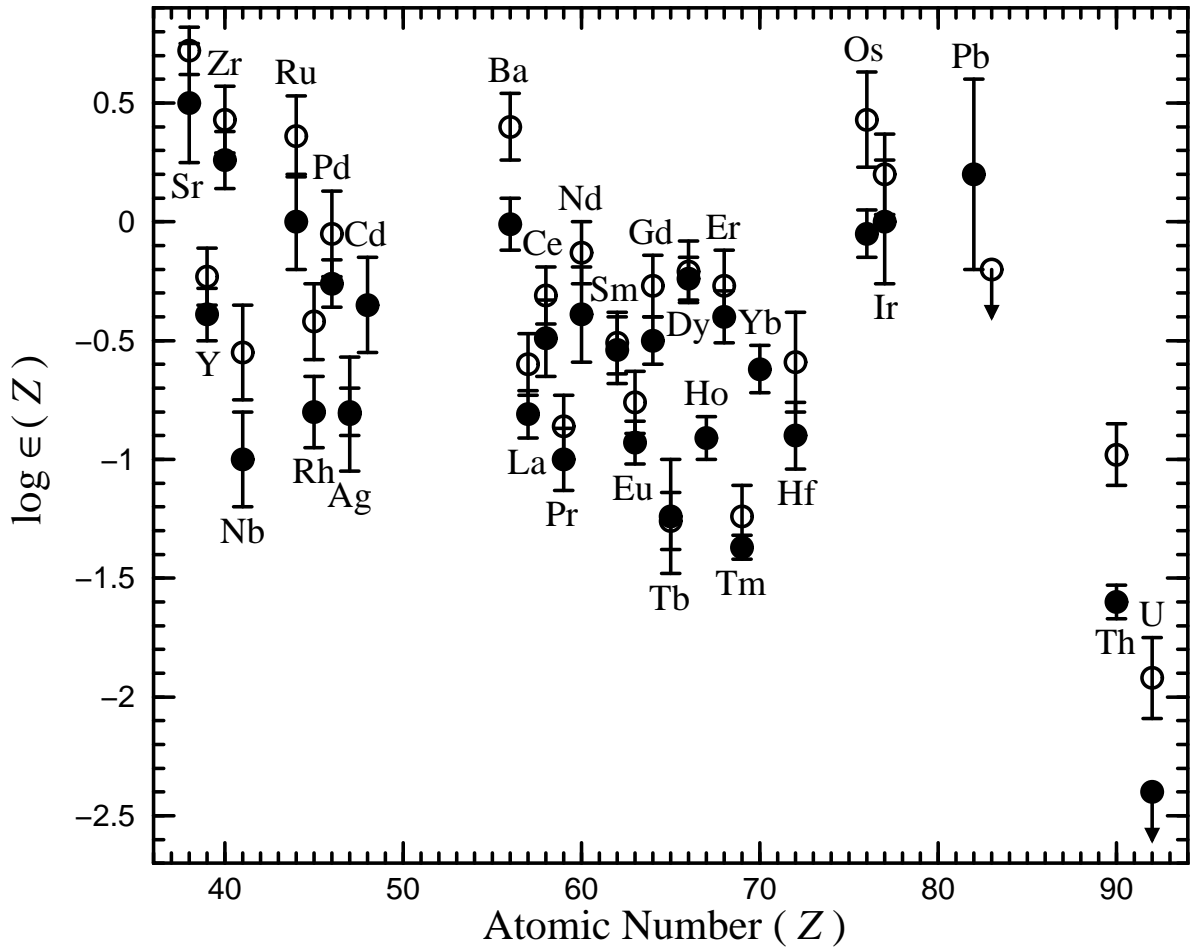}
\end{minipage}
\caption{Comparison of the observed abundances in CS 22892--052 
(filled circles: \cite{sne00}) and CS 31082--001
(open circles: \cite{hil02}). The open circle for Pb is
shifted slightly for clarity. Downward arrows indicate upper limits.}
\end{figure}

However, observations suggest that the actual production of the heavy
$r$-process elements with $A>130$ is quite complex. The data on the
metal-poor stars CS 22892--052 \cite{sne00} and CS 31082--001 \cite{hil02}
are shown in Figure 3. It can be seen that the heavy $r$-process elements
are not produced alone, but along with some light $r$-process elements
such as Sr, Y, and Zr
with $A<130$. Further, while the abundances of the elements below Pb
are essentially identical for the two stars, their Th and U abundances
differ greatly. These two features are in contradiction with the commonly
cited virtues of fission cycling: a robust yield pattern at $A>130$ and
no production of the nuclei with $A<130$ (e.g., \cite{frt99}).

On the other hand, fission cycling may not occur during the $r$-process.
Instead, the progenitor nuclei may fission during decay toward stability.
If the $r$-process initially produces a progenitor pattern covering
nuclei with $A\gtrsim 190$ with a peak at $A\sim 195$, both the light
$r$-process elements with $A<130$ and the heavy ones with $130<A<190$
can be produced by fission of the progenitor nuclei during decay
\cite{qian02}. The fission probability can be significantly enhanced by 
the interaction between supernova neutrinos and the progenitor nuclei
\cite{qian02,kol04}. This is because $\nu_e$ with an average energy of 
$\sim 10$ MeV can excite these nuclei to $\sim 20$ MeV above the
ground state through charged-current interaction while $\nu_\mu$,
$\bar\nu_\mu$, $\nu_\tau$, and $\bar\nu_\tau$ with an average energy of 
$\sim 20$ MeV can provide similar
excitation through neutral-current interaction. Of course, the
extremely neutron-rich progenitor nuclei can also deexcite through
neutron emission. It was shown that the solar $r$-process abundances
at $A=183$--187 can be completely accounted for by neutrino-induced
neutron emission from the progenitor nuclei in the $A\sim 195$ peak
(see Fig. 4) \cite{qian97,hax97}. Interestingly, the level of neutrino 
interaction required to accomplish this will also induce significant 
fission \cite{qian02}.

\begin{figure}[htb]
\begin{minipage}{1.5cm}
\makebox[1cm]{}
\end{minipage}
\begin{minipage}[t]{8cm}
\includegraphics[scale=0.7]{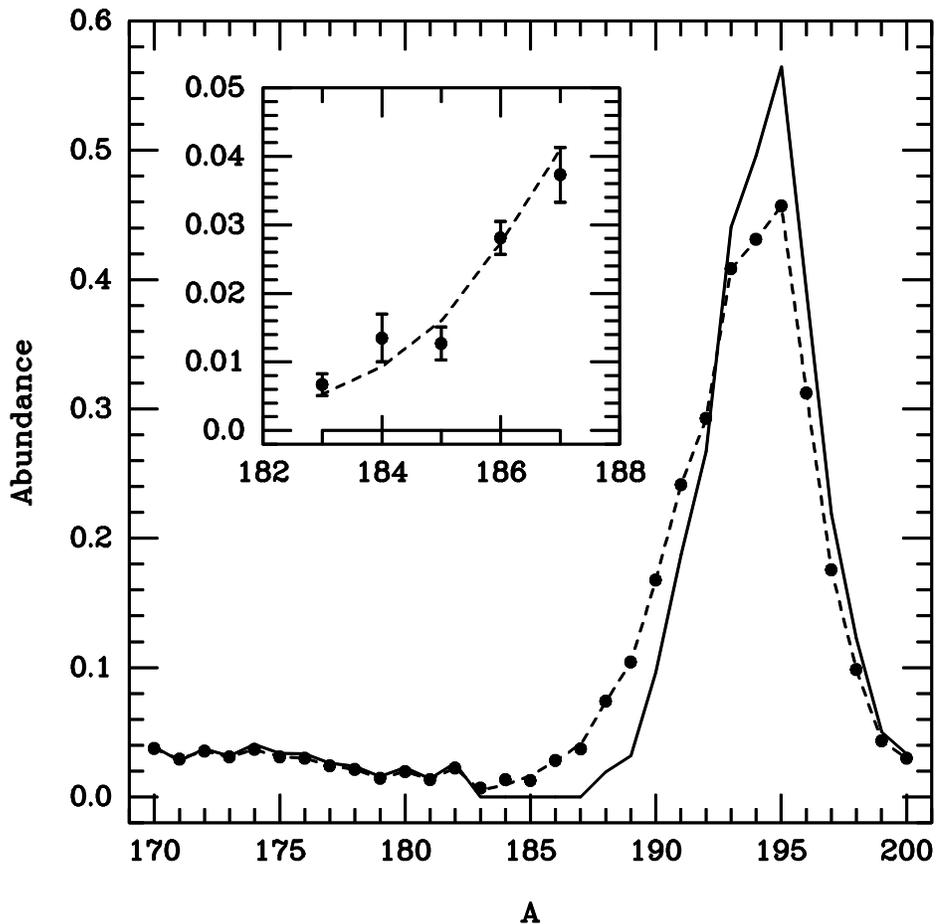}
\end{minipage}
\caption{Production of nuclei by neutrino-induced neutron emission.
The abundances before and after neutrino processing are given by the solid
and dashed curves, respectively. The filled circles (some with error bars)
give the solar $r$-process abundances.}
\end{figure}

\subsection{More details of yield patterns and need for nuclear data}
As can be seen from the above discussion, to understand the details of
the $r$-process patterns requires a great deal of nuclear input regarding
processes such as fission and neutrino interaction. Unfortunately, the
majority of this input has to depend on theoretical calculations.
The danger of this dependence and the importance of experimental
data are illustrated below with an example concerning
neutron separation energies of neutron-rich nuclei far from stability.

\begin{figure}[htb]
\begin{minipage}[t]{8cm}
\includegraphics[scale=0.65,angle=90]{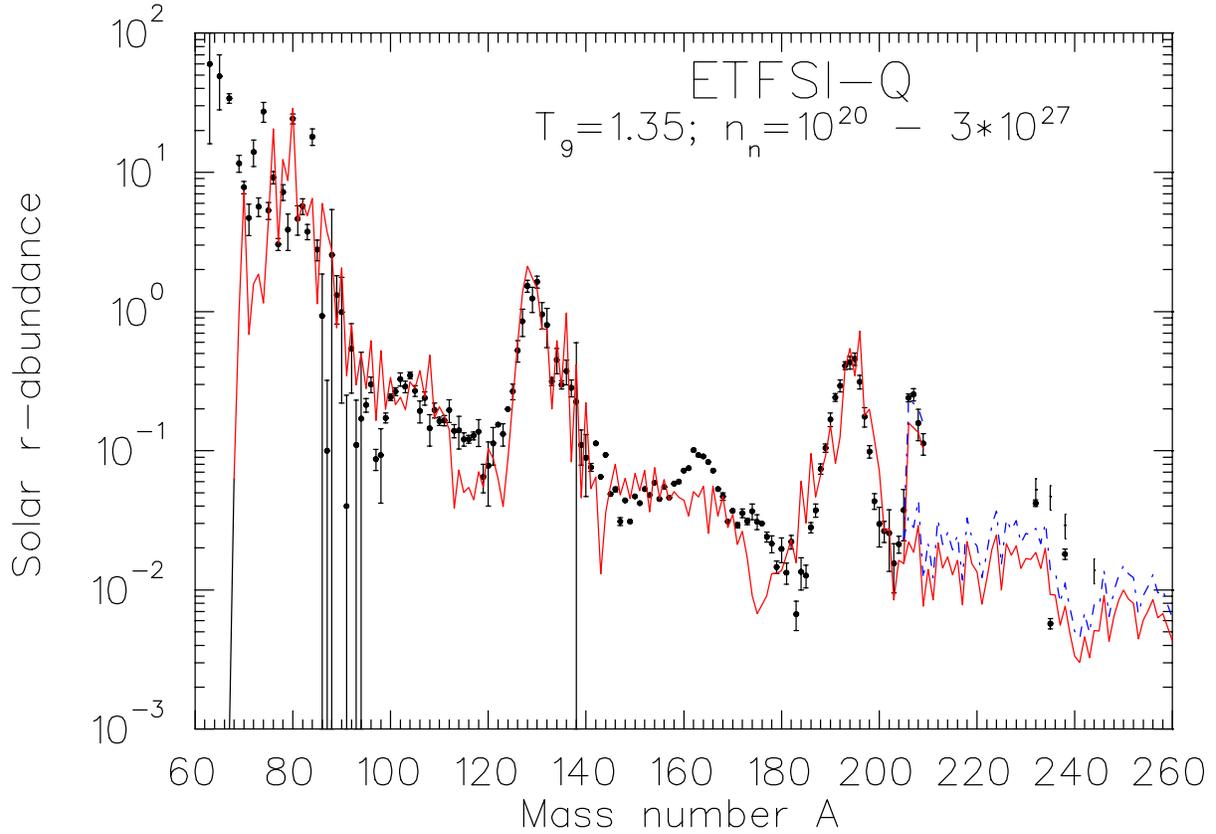}
\end{minipage}
\caption{The solid curve represents the $r$-process pattern calculated in 
\cite{cow99} (the dashed curve has a slightly different weight for $A>206$).
Compared with the solar $r$-process abundances (small filled circles with 
error bars), the calculated pattern has
severe deficiencies below the $A\sim 130$ and 195 peaks.}
\end{figure}

The detailed $r$-process pattern calculated in \cite{cow99} is shown in
Figure 4 along with the solar $r$-process pattern. It can be seen that
the calculated pattern, among other things, greatly underproduces the
nuclei below the $A\sim 130$ and 195 peaks. These deficiencies can be 
traced to the lack of waiting-point nuclei in the corresponding mass 
regions. Due to the tug of war between neutron capture and 
photodisintegration, the waiting-point nuclei have approximately the
same neutron separation energy. In general, at a given neutron number,
the neutron separation energy increases with increasing proton number 
and at a given proton number it decreases with increasing neutron number.
So a nucleus with proton number $Z+1$ must have a larger neutron number
than a nucleus with proton number $Z$ in order to have the same
neutron separation energy. Typical mass formulae used in $r$-process 
calculations predict that for a given proton number
the neutron separation energy decreases very 
slowly as the neutron number increases toward a magic number. 
Consequently, below a closed neutron shell, two nuclei with neighboring 
proton numbers must have a large difference in neutron number in order to 
have the same neutron separation energy. This results in a large mass gap
between two neighboring waiting-point nuclei below the closed neutron 
shell and causes the
severe underproduction below the corresponding abundance peak.
It was shown that the deficiencies discussed above can be alleviated by
quenching the strength of closed neutron shells in the mass formulae
(e.g., \cite{che95}). Clearly, to fully resolve these deficiencies requires
mass measurements in the relevant regions.

\section{CONCLUSIONS}
In summary, observations of abundances in metal-poor stars suggest that
supernovae are the major $r$-process site and there are two different
kinds of $r$-process. Supernovae associated with progenitors of
$\sim 8$--$10\,M_\odot$ and accretion-induced collapse of white dwarfs
are mainly responsible for the heavy $r$-process elements with
$A>130$ while supernovae with progenitors of $>10\,M_\odot$ are mainly
responsible for the light $r$-process elements up to the $A\sim 130$
peak. How the $r$-process conditions are obtained in supernovae is
still quite uncertain. As mentioned above, the neutron-richness of the
neutrino-driven wind is determined by neutrino emission from the
neutron star. Future studies should investigate whether neutron stars
evolved from different presupernova cores have very different neutrino
emission characteristics. The effects of neutrino oscillations should
also be considered (e.g., \cite{qian93}). 
Of course, magnetic fields and rotation, which are
the usual ``villains'' in astrophysical problems, may be important, too.
Further, the $r$-process yield patterns depend on a great deal of nuclear 
input regarding for example, masses, fission, and neutrino responses of 
neutron-rich nuclei far from stability. There are large uncertainties
in the current theoretical nuclear input for the $r$-process calculations. 
Therefore, future measurements are essential in providing either direct
information or guidance to theory. Hopefully, the next generation of
rare isotope accelerator facilities can help put the understanding of 
$r$-process production on a solid basis.

\end{document}